\documentclass[11pt]{article}
\usepackage{graphicx,subfigure,epsfig}
\newcommand{\BABARPubYear}    {01}

\newcommand{\BABARConfNumber} {12}
\newcommand{\SLACPubNumber} {8981}

\input pubboard/babarsym

\newcommand{\pvec}{{\bf p}}
\newcommand{\xf}{\mbox{${\cal F}$}}
\newcommand{\DE}{\ensuremath{\Delta E}}
\newcommand{\sigGSBratio}{\calR}

\long\def\inst#1{\par\nobreak\kern 4pt\nobreak
    {\it #1}\par\vskip 10pt plus 3pt minus 3pt}

\begin{document}
\pagestyle{empty}

\begin{flushright}
\babar-CONF-\BABARPubYear/\BABARConfNumber \\
SLAC-PUB-\SLACPubNumber \\
September, 2001 \\
\end{flushright}


\par\vskip 5cm

\begin{center}
\Large {\bf Measurement of the Branching Fraction for
$B^{+} \ra K^{*0} \pi^{+}$}
\end{center}
\bigskip

\begin{center}
\large The \babar\ Collaboration\\
\mbox{ }\\
\today
\end{center}
\bigskip \bigskip

\begin{center}
\large \bf Abstract
\end{center}

We present a preliminary result of the branching fraction for 
the $B$ meson decay to the final state $K^+\pi^-\pi^+$ via an
intermediate $K^{*0}$ resonance
using the sample of approximately 23 million
\BB\ mesons produced at the $\Upsilon(4S)$
resonance with the \babar\ detector at the PEP II \epem\ collider. 
The $K^{*0}$ was detected through the decay to the final state $K^+ \pi^-$.
The result of this analysis is   
${\cal B}(B^{+} \ra K^{*0} \pi^{+}) = (15.5 \pm 3.4 \pm 1.8) \times 10^{-6}$ 
where the first error is statistical and the second is systematic. 

\vfill
\begin{center}
Submitted to the 
9$^{th}$ International Symposium on Heavy Flavor Physics \\
9/10---9/13/2001, Pasadena, CA, USA
\end{center}

\vspace{1.0cm}
\begin{center}
{\em Stanford Linear Accelerator Center, Stanford University, 
Stanford, CA 94309} \\ \vspace{0.1cm}\hrule\vspace{0.1cm}
Work supported in part by Department of Energy contract DE-AC03-76SF00515.
\end{center}

\newpage
\pagestyle{plain}

\begin{center}
\small

The \babar\ Collaboration,
\bigskip

B.~Aubert,
D.~Boutigny,
J.-M.~Gaillard,
A.~Hicheur,
Y.~Karyotakis,
J.~P.~Lees,
P.~Robbe,
V.~Tisserand
\inst{Laboratoire de Physique des Particules, F-74941 Annecy-le-Vieux, France }
A.~Palano,
A.~Pompili
\inst{Universit\`a di Bari, Dipartimento di Fisica and INFN, I-70126 Bari, Italy }
% ...

\end{center}\newpage

\section{Introduction}
\label{sec:Introduction}

\par Charmless hadronic $B$ meson decays to pseudoscalar-vector final
states have contributions from tree and penguin diagrams that are of
similar magnitudes. It is expected that the interference between the
tree and the penguin contributions may lead to direct $CP$ violation.
The measurement of the branching fractions of these charmless decays 
is an important first step in our understanding of the sizes of the 
various contributions. 
Recent measurements of the decays of $B$
mesons to the pseudoscalar-pseudoscalar states involving pions and
kaons hint at larger penguin contributions than expected \cite{ref:twobody}. 

In this paper we report a preliminary measurement of the branching fraction 
for the decay $B^+ \ra K^{*0} \pi^+$ where the $K^{*0}$ resonance is 
detected through its decay to the $K^+ \pi^-$ final state.
Inclusion of charge conjugate states
is assumed throughout the paper.

\section{The \babar\ Detector and Dataset}
\label{sec:babar}

\par The data used in these analyses were collected with the \babar\
detector at the \pep2\ storage ring ~\cite{ref:pepii}.  
The \babar\ detector, described
in detail elsewhere~\cite{ref:babar}, consists of five active
sub-detectors. Surrounding the beam-pipe is a silicon vertex tracker
(SVT) to track particles of momentum less than $\sim$120\mevc\ and to
provide precision measurements of the positions of charged particles
of all momenta as they leave the interaction point. A beam-support
tube surrounds the SVT. Outside this is a 40-layer drift chamber
(DCH), filled with an 80:20 helium-isobutane gas mixture to minimize
multiple scattering, providing measurements of track momenta in a 1.5
T magnetic field.  It also provides energy-loss measurements that
contribute to charged particle identification. Surrounding the 
drift chamber is a novel detector of internally
reflected Cherenkov radiation (DIRC) that provides charged hadron
identification in the barrel region. The DIRC consists of quartz bars of
refractive index $\sim$1.42 in which Cherenkov light is produced by
relativistic charged particles. The light is internally reflected and
collected by an array of photomultiplier tubes, which enable Cherenkov
rings to be reconstructed and associated with the charged tracks in
the DCH, providing a measurement of particle velocities. Outside the
DIRC is a CsI(Tl) electromagnetic calorimeter (EMC) which is used to
detect photons and neutral hadrons, and to provide electron
identification.  The EMC is surrounded by a superconducting coil which
provides the magnetic field for tracking. Outside the coil, the flux
return is instrumented with resistive plate chambers interspersed with
iron (IFR) for the identification of muons and long-lived neutral hadrons. 

The data sample used for the analyses contains approximately 23~million 
$\BB$ pairs, corresponding to 20.7\invfb\ taken near the
$\FourS$ resonance. An additional 2.6\invfb\ of data were taken
approximately 40\mev
below the \FourS\ resonance to validate the contribution to
backgrounds resulting from \epem\ annihilation into light \qqbar\
pairs.  These data have all been processed with reconstruction
software to determine the three-momenta and positions of charged
tracks and the energies and positions of photons.

\section{Analysis Method}
\label{sec:Analysis}

\subsection{Candidate Selection}

\par The $B$ candidates are reconstructed from three charged tracks.
Charged tracks are required to have  
a transverse momentum greater than $0.1\gevc$, 
at least 12 hits in the DCH and to originate close to the beam-spot.
A good quality vertex of the three tracks is also required which is 96\%
efficient for signal.

One of the tracks is required to pass a tight kaon selection  
based on Cherenkov angle information 
from the DIRC combined with energy-loss information from the DCH.
The track that passes the kaon selection is assigned the kaon mass.
The other two tracks, which are candidate pions, are required to fail the 
kaon selection and are assigned the pion mass.
The kaon selection passes 79\% of signal kaons and  96\%
of signal pions pass the pion identification.

The $B$ candidates are required to satisfy kinematic constraints
appropriate for $B$ mesons produced by the decay of an $\Upsilon(4S)$.  
We use two kinematic variables: the energy-substituted $B$ mass 
 $\mes = \sqrt{(s/2 + \pvec_0\cdot \pvec_B)^2/E_0^2 - p_B^2}$, 
where the subscripts $0$ and
$B$ refer to the \epem\ system and the $B$ candidate, respectively;
and $\Delta E = E_B^* - \sqrt{s}/2$, where $E_B^*$ is the \B\
candidate energy in the center-of-mass frame.  For signal events, the
former has a value close to the $B$ meson mass and the latter should
be close to zero.
In the Monte Carlo simulation of $B^+ \ra K^{*0} \pi^-$ the $\Delta E$ and $m_{ES}$ distributions
are Gaussian and have have resolutions of $18 \mev$ and $2.4\mevcc$ 
respectively.
The requirements on these variables, $\mid \Delta E \mid < 60\mev$ and 
 $\mid m_{ES}-5279.5 \mid < 7.5 \mevcc$, form the signal region
in the $m_{ES}$-$\Delta E$ plane. 
The bounds of the signal region are over three times the Monte Carlo predicted resolutions so that uncertanties in the $\Delta E$ and $m_{ES}$ distributions do not contribute significant systematic errors.

The $K^{*0}$ candidate is formed by adding the 4-momenta of the kaon 
and oppositely charge pion candidates.
A window of $0.816-0.976\gevcc$ is allowed for the mass of the $K^+ \pi^-$
pair, $\pm$1.6 times the width of the $K^{*0}$ resonance. 
For $B$ decays to final states formed from a pseudoscalar and a vector
meson, the vector meson is longitudinally polarized. 
The angular distribution of $K^{*0}$ decay products is used to suppress background.
The helicity angle, $\theta_H$, is defined as the angle of the $K$ momentum 
in the $K^{*0}$ rest frame with respect to the momentum of the $K^{*0}$ in the 
lab frame.
The distribution of $\theta_H$ follows a $\cos^2 \theta_H$ functional form for signal and is essentially flat for background.
A requirement of $\mid \cos{\theta_H} \mid > 0.2$ is used in this analysis.

To avoid bias, the on-resonance events in the signal region were not counted 
or studied until the analysis method and selection criteria were finalised.
Many of the selection criteria were optimized for statistical significance 
to an assumed branching fraction, as the signal region was blinded. 
The criteria that were optimized are the mass window on the  
 $K^*$ candidates, $\cos{\theta_H}$, the criteria on the particle 
identification and the requirements on event shape
discussed in the following section.

\subsection{Background Suppression and Characterization}
\label{sec:background}

\par There are two contributions to the background. The dominant background is
random combinatorial background but there is also a small contribution 
from specific $B$ decay channels.

Charmless hadronic modes suffer large amounts
of background from random combinations of tracks, mostly from light quark
and charm continuum production. Such backgrounds may be reduced by 
selection requirements on the event topologies computed in the 
 $\FourS$ rest frame. 
As the $B$ mesons are almost at rest in this frame, they decay spherically, 
whereas continuum events are jet--like. 
We use $\cos{\theta_T}$, the cosine of the angle 
$\theta_T$ between the thrust axis of the $B$ meson decay and the 
thrust axis of the rest of the event (all tracks and neutral clusters in 
the event except those that form the $B$ candidate). 
For continuum-related backgrounds, 
these two directions tend to be aligned because the fake reconstructed 
$B$ candidate daughters lie in the same jets as the remaining particles
in the event. By contrast, in $B$ events, the decay 
products from one $B$ meson are independent of those in the other, 
making the distribution of this angle isotropic. 
A requirement on the maximum size of $|\cos{\theta_T}|$ strongly suppresses continuum background. 
This requirement was optimised to give the greatest significance to the branching fraction.
 $\mid \cos{\theta_T} \mid < 0.7$ was optimum which is 67\% efficient for signal and rejects 90\% of continuum events.

Other event shape variables also help to discriminate between $b \overline{b}$ 
and continuum events. 
We form a linear combination of 11 variables in a Fisher discriminant 
\xf\ \cite{Fisher}.
The coefficients for each variable are chosen to maximize the
separation between training samples of signal and background events. 
The variables contained in \xf\ are
the cosine of the angle between the $B$ momentum and the beam axis,
the cosine of the angle between the thrust axis of the $B$ candidate
and the beam axis, and the summed momentum of the rest of the event
in nine cones coaxial with the thrust axis of the $B$ candidate.
The \xf\ requirement is 80\% efficient on signal and rejects 60\% of continuum events
after the $\cos{\theta_T}$ cut has been applied.

Despite the power of such topological variables to reduce the combinatorial 
backgrounds, it is necessary to make a background subtraction in measuring 
the branching fraction. In order to do this, the background in the 
signal region is estimated from the number of on-resonance data events 
in a sideband region and extrapolated into the signal region. 
The sideband region is a larger area than the signal region at lower $\mes$ 
and is specified by $5.22 < \mes < 5.26\gevcc$, $|\DE|<0.3\gev$.
In order to extrapolate from the sideband region to the signal region the shape of the background in the $\Delta E$-$m_{ES}$ plane must be characterised.
This is done using on-resonance data (avoiding the signal region), off-resonance data and Monte Carlo continuum events in the region $5.20<m_{ES}<5.29\gevcc$, $|\Delta E|<-0.3\gev$.
The shape of the distribution of the background as 
a function of \mes\ is parameterized using the ARGUS function 
\cite{ref:argus} which has one shape parameter, $\xi$.
The background $\Delta E$ distribution is quadratic and the integrals over 
the signal region and sideband depend only on the quadratic coefficient.
The  quadratic coefficient and $\xi$ values measured for on- and off-resonance data and continuum Monte Carlo (which are consistent) are averaged and
used along with the $\Delta E$ and $m_{ES}$ requirements to 
calculate \sigGSBratio, the ratio of the number of candidates in the 
signal region to the number in the sideband region. 
For this analysis, $\sigGSBratio =  0.0647\pm 0.00029$.

In order not to underestimate the background, the selection criteria optimisation is 
done using only half the on-resonance events (every alternative event) and 
the final number of events in the sideband is measured on the other half of 
the events.

The $B$ background is supressed by the particle identification,
and the $K^*$, $\Delta E$, and $m_{ES}$ requirements.
The remaining $B$ background is then estimated with Monte Carlo events and 
subtracted from the number of events in the signal region.
Possible sources of $B$-related backgrounds include events with
low-multiplicity decays to charm and other charmless decays. 
Background channels were identified by running over generic 
$b\bar{b}$ Monte Carlo and  a dedicated charmless $B$ decay 
simulation with loosened analysis requirements.
For each background, Monte Carlo events exclusively of that type were studied 
to predict feed-through. 
Current branching fractions \cite{ref:pdg} are used to calculate the 
number of events expected in the signal region.
Three channels have potential contributions, $B^+ \ra K^{*0} K$, $B^+ \ra \rho^0 K^+$ and $B^+ \ra \rho^0 \pi^+$ though they are all small.
The total amount of $B$-related background expected in the signal region is 
only $1.0\pm 0.6$ events, the error is from the errors on the branching 
fractions and is systematic. 

Non-resonant $B^+ \ra K^{+}\pi^{-}\pi^{+}$ decays and decays to a $K^+ \pi^- \pi^+$ final states proceeding through higher $K^+ \pi^-$ resonances are expected to be negligable in the signal region (less than one event).
Although backgrounds from channels with the same final state are small the possible interference of the amplitudes of these processes may affect the branching fraction in the area of the Dalitz plot under study.
This effect is not considered in the the branching fraction calculation but could be as large as 30\%.
With more data this interference can be studied.

\subsection{Branching Fraction Analysis}
\label{sec:Physics}

\par The branching fraction is calculated using 
\begin{equation}
{\cal B} = \frac{N_1-2\sigGSBratio N_2-N_{Bbg}}{N_{\BB} \times \epsilon}
\label{BReqn}
\end{equation}
where $N_1$ is the number of candidates in the signal region for
on-resonance data; $N_2$ is the number of candidates in half the 
on-resonance data 
observed in the sideband region, so that $2\sigGSBratio N_2$ is the 
estimated number of background candidates in the signal region;
$N_{Bbg}$ is the number of  $B$-related background expected in the signal region;
$N_{\BB}$ is the number of $\BB$ pairs produced and $\epsilon$ is the signal 
efficiency multiplied by ${\cal B}(K^{*0} \ra K^+ \pi^-) = 2/3$.

For the signal efficiency in Eq.~\ref{BReqn}, we used simulated signal
events, applied the same selection criteria as used for data, and corrected 
for discrepancies between Monte Carlo and data. 
The efficiencies due to tracking, particle identification, vertex 
quality and \xf\ selection criteria are determined from independent 
control samples derived from the data.

Any discrepancy between the measured track-finding efficiency 
and that obtained in our simulation is taken into account with a 
momentum, azimuthal angle, and track multiplicity dependent weight per track. 
The $\pi^{\pm}$ and $K^{\pm}$ particle identification efficiency was
measured in data using a control sample of $D^+ \ra \overline{D}^0\pi^+, 
\overline{D}^0 \ra K^+\pi^-$.  The efficiency was measured as a function 
of particle momentum and polar and azimuthal angle.

The efficiency of the \xf\ requirement was measured using the control sample 
 $B^+ \ra \overline{D}^0\pi^+, \overline{D}^0 \ra K^+\pi^-$ which has
the same final state and similar kinematics to the signal mode.
The vertex quality requirement efficiency was not measured on this channel
as the $D^0$ may fly before decaying.
Instead the control sample $B^+ \ra J/\psi K^+, J/\psi \ra \mu^+ \mu^-$
was used.

\section{Treatment of Experimental Uncertainties}
\label{sec:Systematics}

\par The systematic errors associated with branching fraction measurement
arise from uncertainties in the background subtraction, in the overall
signal efficiency and in the number of \BB pairs produced as this is only 
known to 1.6\%.

The estimate of the combinatoric background is given as the product of 
the number of events in the sideband region and the factor \sigGSBratio.
The number of events observed in the sideband has only a statistical error.
The factor \sigGSBratio\ has a systematic error given by the error in the
fitted ARGUS shape parameter, $\xi$. The contribution from the error 
in the $\Delta E$ characterization is negligible. 
A systematic uncertainty on \sigGSBratio\ is estimated
from the range of values of $\xi$ measured 
in different data and different $\Delta E$ regions.
The error on \sigGSBratio\ contributes a 2.5\% systematic uncertainty to the branching fraction.
The error on $N_{Bbg}$ the number of $B$-related events expected in the signal
 region, is from the errors on the branching ratios used in the calculation.
This contributes 1.8\% to the systematic error.

Some uncertainty arises from the limited statistics of the signal Monte Carlo
used to estimate the efficiency.
The fractional error from this source is 2.3\%.
The accuracy of the simulation is subject to
systematic uncertainties leading to uncertainties in the efficiencies of 
selection requirements. At present the dominant uncertainties are in the
the efficiencies of the  particle identification (6.4\%), vertex quality cut (6.0\%), \xf\ cut (4.1\%) and tracking (3.6\%).
These systematic errors come from the methods used to to calculate the 
efficiencies in data, explain above.

The uncertainty on the particle identification data efficiency is taken as 
5.0\% per kaon and 2.0\% per pion.  The particle identification  efficiency
errors for the two pions are added coherently but the pion and kaon 
identification efficiencies have 
little correlation and so are added in quadrature.  
The uncertainty in the determination of the tracking efficiency in data 
leads to an uncertainty in the tracking efficiency of 1.2\% 
per track, added coherently for all charged tracks in the $B$ candidate. 

The overall systematic uncertainty is the sum in quadrature of the 
contributions from all sources giving a systematic error
on the branching fraction of 11\%.

\section{Physics Results and Conclusion}
\label{sec:results}

\par
\begin{figure}
\begin{center}
\begin{tabular}{c} 
  \epsfig{file=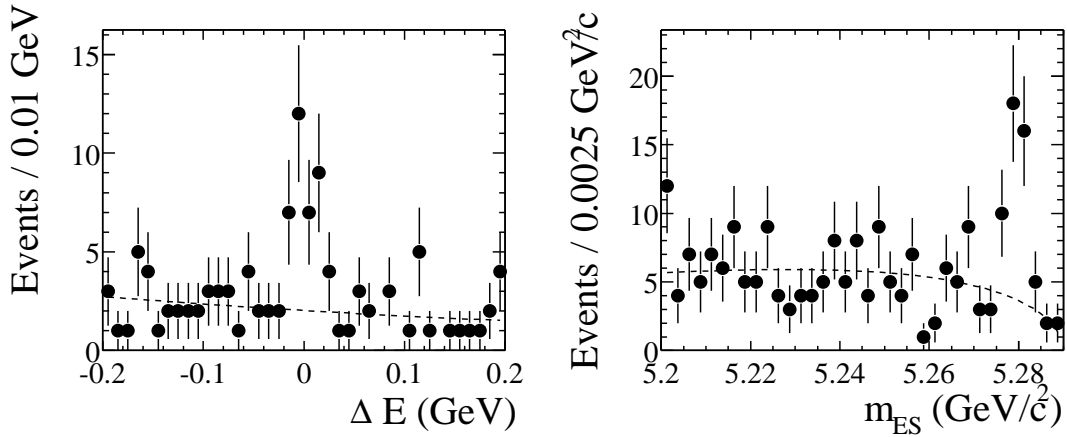,width=0.9\linewidth}
\end{tabular}
\caption{$\Delta E$ and $m_{ES}$ distributions for 
$B^{+} \ra K^{*0} \pi^{+}, K^{*0} \ra K^+\pi^-$. The lines are the background distributions used in the background subtraction.}
\label{fig-results}
\end{center}
\end{figure}

\begin{figure}
\begin{center}
\begin{tabular}{c} 
  \epsfig{file=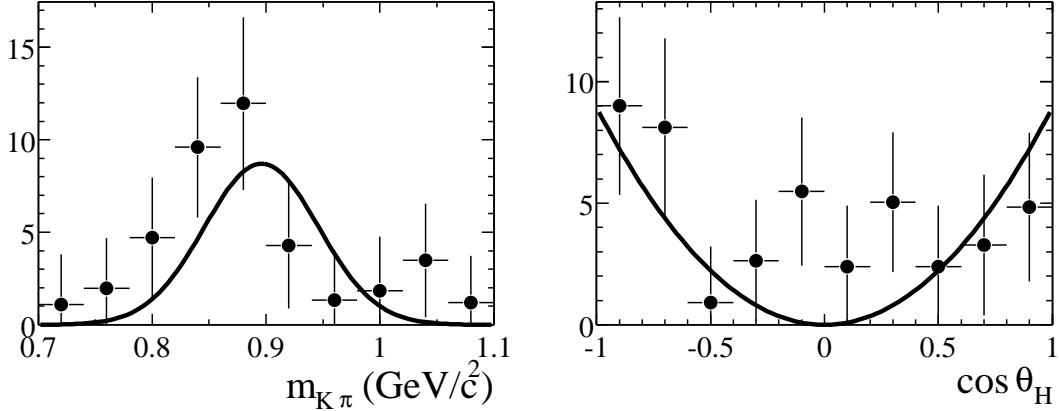,width=0.9\linewidth}
\end{tabular}
\caption{The mass of the $K^{*0}$ candidate and $\cos\theta_H$ for on-resonance data in the signal region with background subtracted. The lines are the expexted $B^+ \ra K^{*0} \pi^+$ distributions.}
\label{fig-kstar}
\end{center}
\end{figure}

 Our preliminary measurement result of the branching fraction is
summarized in Table~\ref{tab-results}.
Figure~\ref{fig-results} shows the $\Delta E$ and $m_{ES}$
distributions in the signal region with lines showing the background distributions used in the background subtraction.
Figure~\ref{fig-kstar} shows the background subtracted $K^{*0}$ candidate mass
and $\cos\theta_H$. The lines on the plots are the expected $B^+ \ra K^{*0} \pi^+$ distributions.
The data is consistent with being polarised $K^{*0}$ mesons as the $\chi^2$ per degree of freedom is less than unity for both distributions.

\begin{table}[ht]
\begin{center}
{\small
\begin{tabular}{|l|c|}
\hline
Events in signal region &     55 \\
Events in sideband      &     148 \\
Ratio, $\cal R$         &     $0.0647\pm 0.0029$ \\
Combinatoric background &     $19.1 \pm 1.6 \pm 0.9$ \\
$B$-related background  &     $1.0 \pm 0.6$ \\
signal yield            &    $34.8 \pm 7.6 \pm 1.1$ \\
signal efficiency       &     $(0.149 \pm 0.004 \pm 0.013) \times 2/3$ \\
\hline
Stat. Signif. ($\sigma$) & 6.0 \\
$\cal B$ ($\times 10^{-6}$) & $15.5\pm3.4\pm1.8$\\
\hline
\end{tabular}
}
\end{center}
\caption
{A summary of the branching fraction measurement analysis for the decay mode $B^{+} \ra K^{*0} \pi^{+}$. The factor of $2/3$ in the signal efficiency is the branching fraction of $K^{*0}\ra K^+ \pi^-$.}
\label{tab-results}
\end{table}


In summary, we have measured the branching fraction for $B$
meson decay to the final state $K^+\pi^-\pi^+$ via an
intermediate $K^{*0}$ resonance with a statistical significance of $6.0\sigma$.
Our preliminary result is
 ${\cal B}(B^{+} \ra K^{*0} \pi^{+}) = (15.5 \pm 3.4 \pm 1.8) \times 10^{-6}$.


\section{Acknowledgments}
\label{sec:Acknowledgments}
We are grateful for the 
extraordinary contributions of our \pep2\ colleagues in
achieving the excellent luminosity and machine conditions
that have made this work possible.
The collaborating institutions wish to thank 
SLAC for its support and the kind hospitality extended to them. 
This work is supported by the
US Department of Energy
and National Science Foundation, the
Natural Sciences and Engineering Research Council (Canada),
Institute of High Energy Physics (China), the
Commissariat \`a l'Energie Atomique and
Institut National de Physique Nucl\'eaire et de Physique des Particules
(France), the
Bundesministerium f\"ur Bildung und Forschung
(Germany), the
Istituto Nazionale di Fisica Nucleare (Italy),
the Research Council of Norway, the
Ministry of Science and Technology of the Russian Federation, and the
Particle Physics and Astronomy Research Council (United Kingdom). 
Individuals have received support from the Swiss 
National Science Foundation, the A. P. Sloan Foundation, 
the Research Corporation,
and the Alexander von Humboldt Foundation.


\begin{thebibliography}{99}
\label{sec:biblio}



\bibitem{ref:twobody} 
\babar\ Collab., B. Aubert et al, hep-ex/0105061. Belle Collab., K. Abe et
al, hep-ex/0104030. CLEO Collab., D. Cronin-Hennessy et al, PRL 85, 515
(2000).





\bibitem{ref:pepii}
PEP-II Conceptual Design Report, SLAC-R-418 (1993).

\bibitem{ref:babar}
\babar\ Collaboration, B.\ Aubert {\em et al.},
SLAC-PUB-8596,
to appear in Nucl.\ Instrum.\ Methods.




\bibitem{Fisher} CLEO Collaboration, D.\ M.\ Asner {\it et al.}, 
Phys.~Rev.~{\bf D53}, 1039 (1996).

\bibitem{ref:argus}
ARGUS Collaboration, H.~Albrecht {\it et al.}, \jpl {\bf B185}, 218 (1987).

\bibitem{ref:pdg} Particle Data Group, D.~E.~Groom {\it et al.},
\epjc {\bf 15}, 1 (2000).



\end{thebibliography}
\end{document}